\documentstyle[12pt]{article}

\def\be{\begin{equation}}
\def\ee{\end{equation}}
\def\ben{\begin{displaymath}}
\def\een{\end{displaymath}}
\def\ba{\begin{array}{c}}
\def\bal{\begin{array}{l}}
\def\ea{\end{array}}
\def\p{\partial}
\begin{document}

%.

\vspace{1.5cm}

 \begin{center}{\Large \bf

Determination of the domain of the admissible matrix elements in
the four-dimensional ${\cal PT}-$symmetric anharmonic model

  }\end{center}

\vspace{10mm}

 \begin{center}

 {\bf Miloslav Znojil}

 \vspace{3mm}
\'{U}stav jadern\'e fyziky AV \v{C}R,

250 68 \v{R}e\v{z},

Czech Republic

{e-mail: znojil@ujf.cas.cz}

\vspace{3mm}

\vspace{5mm}

 % \today, four.tex

\end{center}

\vspace{5mm}

\section*{Abstract}

Many manifestly non-Hermitian Hamiltonians (typically, ${\cal
PT}-$symmetric complex anharmonic oscillators) possess a strictly
real, ``physical" bound-state spectrum. This means that they are
(quasi-)Hermitian with respect to a suitable non-standard metric
$\Theta \neq I$. The domain ${\cal D}$ of the existence of this
metric is studied here for a nontrivial though still non-numerical
four-parametric ``benchmark" matrix model.

\newpage

\section{Non-Hermitian observables with real spectra
\label{locals}}

Quite a few realistic quantum models are characterized by a mere
``fragile" stability of their bound states. For example,
the reality of the energies of certain nuclear-physics models
may be lost after an ``inessential" change of its coupling strengths
\cite{Geyer}. Such a phenomenon is rendered possible when the Hamiltonian merely remains Hermitian
with respect to a nonstandard scalar product, i.e., with respect
to a ``nontrivial metric" $\Theta\neq I$ in an underlying
``physical" Hilbert space. This means that all the operators of
observables  $H$ must obey an unusual rule
 \be
 H^\dagger= \Theta\,H\,\Theta^{-1}\neq H\,,\ \ \ \ \ \ \
 \Theta=\Theta^\dagger >0\,.
 \label{prva}
 \label{quasih}
 \label{quasihermi}
 \ee
Such a property of (typically: Hamiltonian) $H$ guarantees the
reality of the spectrum and may be called quasi-Hermiticity.

Practical objections may occur against the use of a nontrivial
metric $\Theta \neq I$. One of their sources lies in the fact that
many phenomenological models in Quantum Mechanics are based on a
{\em differential-operator} realization of their Hamiltonians.
Thus, whenever one works with a one-dimensional
differential-operator Hamiltonian (in units $\hbar = 2m = 1$),
 \be
 H = -\frac{d^2}{dx^2} + V(x) = H^\dagger,
 \ \ \ \ \ \
 x \in (-\infty,\infty)\,
 \label{hermi}
 \ee
one usually prefers the most economical $\Theta=I$ scenario since
even its simplest alternatives require a number of additional
mathematical considerations \cite{Francesco}. Moreover, up to the
very recent past it has been intuitively expected that the
transition to any non-Hermitian generalization of the class of
differential-operator Hamiltonians (\ref{hermi}) would be
accompanied by a complexification of their spectrum and by
a decaying-state re-interpretation of the corresponding wave functions.
The latter misunderstanding
even survived the publication  of a few studies \cite{Caliceti,BG}
which paid attention to specific examples
 \be
 H = -\frac{d^2}{dx^2} + U(x) + {\rm i}\,W(x) \neq H^\dagger,
 \ \ \ \ \ \
 x \in (-\infty,\infty)\,.
  \label{ptsp}
 \ee
where the {\em two independent} real potentials $U(x)=U(-x)$ and
$W(x) = -W(-x)$ happened to generate the {\em real} bound-state
spectra \cite{DDT} and where their spatial symmetry/antisymmetry
has been re-interpreted as a combined parity-plus-time-reversal
(called ``${\cal PT}$") symmetry of the Hamiltonian \cite{DB}.

The scepticism (well sampled, e.g., on the Streater's webpage
\cite{RFS}) did not fully die out even after publication of
several analytic, semi-classical and numerical studies of some new
${\cal PT}-$symmetric models (\ref{ptsp}) in 1998 (cf., e.g.,
\cite{BB,CJT,FG}). Their authors demonstrated that the {\em entire
spectrum} seems to remain {\em real} in comparatively large
domains ${\cal D}$ of parameters. At present, fortunately, we
witness a reconciliation and final acceptance of the idea that the
{\em complex} differential Hamiltonians may possess the {\em real}
bound-state spectra, indeed. The two sets of the fresh 2006
``state-of-the-art" reports may be found in the August dedicated
issue of J.~Phys.~A: Math.~Gen.~(vol.~39, number 32, pp.~9963 --
10261) and in the September dedicated issue of
Czech.~J.~Phys.~(vol.~56, number 9, pp.~885 -- 1064).

One of the paradoxes accompanying such a development of the
subject is that in contrast to the popularity of the various
anharmonic-type models (based, first of all, on their high
relevance in field theory \cite{BM}), much less attention has been
paid to the finite-dimensional matrix versions of the
non-Hermitian quantum Hamiltonians \cite{Wang}. One of the reasons
is that in the matrix models (encountered, quite naturally, in
variational calculations \cite{Geyer}) one usually deals with
``too many" independent matrix elements. The selection and/or
preference of some of them might look ``too arbitrary" in the
context of physics and/or ``too ambiguous" in the language of
mathematics.

In what follows we intend to fill the gap and to study a model
which tries to circumvent both these ``traps" by containing just a
few ``relevant" free parameters {\em and} by being still
surprisingly rich in its mathematical structure and consequences.
Moreover, its ``derivation" from the differential operator
(\ref{ptsp}) (cf. section \ref{ctyrikratctyri}) gives it a certain
generic character while leaving it still purely non-numerical and
exactly solvable. Last but not least, the not quite expected
closed-form feasibility of its mathematical analysis (cf. sections
\ref{analysis} and \ref{Margot}) is quite well matched by some of
its appealing phenomenological features, a few remarks on which
are added here in our final section~\ref{dodatekkectyrem}.

\section{Matrix toy  models
 \label{ctyrikratctyri}}

\subsection{Variational origin}

In order to interconnect the differential and finite-dimensional
$N-$state Hamiltonians  let us start from their most elementary
differential harmonic-oscillator example $H_0$ with the
eigenstates $|\,n\rangle, n=0,1,2,\ldots$ or, in the coordinate
representation,
 \be
 \langle x|\,2m\rangle = {\cal N}_{(m,+)}\,e^{x^2/2} {\cal H}_{2m}(x^2),
 \ \ \ \ \ \ \
 \langle x|\,2m+1\rangle =x\, {\cal N}_{(m,-)}\,e^{x^2/2}
  {\cal H}_{2m+1}(x^2)\,
  \label{normy}
 \ee
where the symbols $ {\cal H}_n$ denote the well known Hermite
polynomials \cite{Fluegge}. The subscripts $\pm$ in the
normalization factors ${\cal N}_{(m,\pm)}$ are added to emphasize
that our basis states (\ref{normy}) are, simultaneously,
eigenstates of the operator of parity ${\cal P}$ with eigenvalues
$(-1)^{n}$. The action of the complex conjugation ${\cal T}$
(mimicking the time reversal \cite{BB}) preserves these basis
states once the normalization factors ${\cal N}_{(m,\pm)}$ are
chosen real.

The ${\cal PT}-$symmetry $H {\cal PT} = {\cal PT} H$ of a given
Hamiltonian (say, (\ref{ptsp})) with real spectrum enables us to
normalize all the eigenstates $|\,\psi_n\rangle$ of $H$ in such a
way that
 \be
 {\cal PT}\,\langle x|\,\psi_n\rangle = + \langle x|\,\psi_n\rangle\,.
 \label{unbroken}
 \ee
In effect, the fixed parity of our harmonic-oscillator
basis~(\ref{normy}) is generalized to the ${\cal PT}-$symmetry of
eigenstates. Once we accept such a convention (speaking about
``unbroken ${\cal PT}-$symmetry of wave functions") we may
decompose
 \be
  |\,\psi\rangle
 =|\,\psi_+\rangle -{\rm i}\,|\,\psi_-\rangle
 \label{rozklad}
 \ee
with  {\em real} expansion coefficients $\phi_m$ and $\chi_m$ in
the variational ansatz of the form
 \ben
 |\,\psi_+\rangle=
 \sum_{m=0}^{N_+}\,|\,2m\rangle \,\phi_m , \ \ \ \ \ \ \
 |\,\psi_-\rangle=
  \sum_{m=0}^{N_-}\,|\,2m+1\rangle \,\chi_m, \ \ \ \ \ \ \
  N_+ + N_- = N \to \infty\,.
  \een
The partitioning of our basis with $|\,2n\rangle \equiv
|\,n_+\rangle$ and $|\,2m+1\rangle \equiv |\,m_-\rangle$ and its
variational truncation with $N \gg 1$ transform Hamiltonian
(\ref{ptsp}) into a finite-dimensional partitioned complex matrix
 \ben
 \tilde{H} = \left (
 \begin{array}{cc}
 S&{\rm i}\,B\\
 {\rm i}\,C&L
 \ea
 \right )
 \een
where the untilded letters denote the submatrices with  {\em real}
matrix elements,
 \be
 S_{mn}=
 \left \langle 2m \left |\,-\frac{d^2}{dx^2} + U(x)\,\right |\,2n
 \right \rangle =
  \left ( S^T \right )_{mn},
  \label{eleraz}
  \ee
  \be
 L_{mn}=
 \left \langle 2m+1 \left |\,-\frac{d^2}{dx^2} + U(x)\,
 \right | \,2n+1 \right \rangle =
  \left ( L^T \right )_{mn},
  \label{eledva}
  \ee
 \be
 B_{mn}=
 \langle 2m|\,W(x)\,|\,2n+1\rangle  \neq
  \left ( B^T \right )_{mn} \equiv C_{mn}
 =
 \langle 2m+1|\,W(x)\,|\,2n\rangle\,.
  \label{eletri}
  \ee
The superscript $^T$ denotes transposition.

\subsection{Real matrix  Schr\"{o}dinger equations}

After we insert (\ref{rozklad}) in Schr\"{o}dinger equation
$\tilde{H}\,|\,\psi\rangle = E\,|\,\psi\rangle $ we reveal that
the resulting partitioned  matrix form of Schr\"{o}dinger equation
for bound states is  real and non-Hermitian,
 \be
  H\,
 \left (
 \begin{array}{c}
 \vec{\phi}\\
 \vec{\chi}
 \ea
 \right )= E\,
 \left (
 \begin{array}{c}
 \vec{\phi}\\
 \vec{\chi}
 \ea
 \right )\,,
 \ \ \ \ \ \ \
 H=
 \left (
 \begin{array}{cc}
 S&B\\
 -B^T&L
 \ea
 \right )
 \,.
 \label{prob}
 \ee
No mathematical contradiction appears in the latter picture since
the metric $\Theta$ naturally becomes singular on the boundary $\p
{\cal D}$ of the domain ${\cal D}$.

In general, the solutions of eq.~(\ref{prob}) must be constructed
numerically. The well known exception is represented by the
two-state models \cite{Turek}. The two-dimensional version of our
present simplified eq.~(\ref{prob}) has also thoroughly been
discussed in our recent letter \cite{Hendrik}. In the
corresponding two by two matrix Schr\"{o}dinger equation
 \be
 \left (
 \begin{array}{cc}
 s&b\\
 -b&l
 \ea
 \right )\,
 \left (
 \begin{array}{c}
 {\phi}\\
 {\chi}
 \ea
 \right )= E\,
 \left (
 \begin{array}{c}
 {\phi}\\
  {\chi}
 \ea
 \right )\,
 \label{prob2}
 \ee
parameters $s,b$ and $l$ are real and, by assumption, the even and
odd unperturbed energies are nondegenerate, $s \neq l$. Via a
suitable scaling we may achieve that $l-s=2$. A shift of the
energy scale $ E \to E + const$ leads to the completely symmetric
arrangement of our entirely general $H$ with $s=-1$ and $l=1$. We
recall the secular equation $\det (H-E)=0$ and deduce the energy
levels,
 \be
 E=E_{\pm}=\pm \sqrt{1-b^2}\,.
 \ee
Thus, for our single-parametric family $H=H(b)$ of the $N=2$
Hamiltonians the domain ${\cal D}={\cal D}(N)$ of the (single)
free parameter $b$ where the energies are real coincides with the
(closed) interval of $b \in {\cal D}(2) \equiv [-1,1]$. At both
the ends of this interval our Hamiltonian ceases to be
diagonalizable. For this reason the domain of the
quasi-Hermiticity of $H$ is often being re-defined as a mere open
set or interior ${\cal D}^{(0)}(2)=(-1,1)$. Under both these
conventions, one finds complex energies in the vicinity of
every element of the boundary $\p {\cal D}\equiv \p
{\cal D}^{(0)}$ \cite{Kato}.

\subsection{Anharmonic-oscillator-like four-by-four matrix
model}

In the harmonic-oscillator model itself the evaluation of the
matrix elements remains trivial and one arrives at the simplest
illustrative example $H_0$ containing just a decoupled pair of
diagonal submatrices,
 \be
 S_{mn}^{(0)}=
 \left \langle 2m \left |\,H_0\,\right |\,2n
 \right \rangle = \delta_{mn}\cdot (4n+1),
 \label{hoca}
 \ee
 \be
 L_{mn}^{(0)}=
 \left \langle 2m+1 \left |\,H_0\,\right |\,2n+1
 \right \rangle = \delta_{mn}\cdot (4n+3).
 \label{hocb}
 \ee
For all the Hermitian generalizations of $H_0$ with unbroken
parity (${\cal P}\,H=H\,{\cal P}$) equation (\ref{prob}) would
stay decoupled ($B=B^T=0$). This means that the parity-preserving
and Hermitian anharmonicities may be considered ``trivial" in
leaving the matrices $S$ and $L$ decoupled and diagonalizable by
the separate unitary transformations in the respective even-parity
and odd-parity subspaces.

We intend to employ just the diagonalized and purely harmonic
submatrices (\ref{hoca}) and (\ref{hocb}), studying merely the
role of the off-diagonal anharmonic-oscillator-like coupling
matrices $B$ in what follows. Thus, we shall start from the
general ${\cal PT}-$symmetric model (\ref{prob}) with
  \be
 H  = \left (
 \begin{array}{ccc|ccc}
 1&0&\ldots&  {}{B}_{11}&{}{B}_{12}&\ldots\\
 0&5&\ddots&{}{B}_{21}&\ldots&\\
 \vdots&\ddots&\ddots&\vdots&&\\
 \hline
 -{}{B}_{11}&-{}{B}_{21}&\ldots&3&0&\ldots\\
 -{}{B}_{12}&\ldots&&0&7&\ddots\\
  \vdots&&&\vdots&\ddots&\ddots
 \ea
 \right )\,.
 \label{nehermaticc}
  \ee
We shall restrict our attention to the ``first nontrivial"
four-level system with the truncated dimensions $N_+=N_-=2$. For
the sake of convenience we shall also symmetrize the unperturbed
spectrum via a shift of the origin on the energy scale, $(1,3,5,7)
\to (-3,-1,1,3)$ and arrive at the Schr\"{o}dinger-equation
 \be
 \left (\begin {array}{cc|cc}
 -3&0&c&b\\
  0&1&a&d \\
  \hline
 -c&-a&-1&0\\
 -b&-d&0&3
 \end {array}
 \right )\,
 \left (
 \begin{array}{c}
 {\phi_0}\\
 {\phi}_1\\
 \hline
 {\chi_0} \\ \chi_1
 \ea
 \right )= E\,
 \left (
 \begin{array}{c}
 {\phi_0}\\
 {\phi}_1\\
 \hline
 {\chi_0} \\
 \chi_1
 \ea
 \right )\,.
 \label{prob4}
 \ee
In comparison with the current two-state analyses, a combined
effect mediated by the {\em simultaneous} growth of all the four
real parameters $a$, $b$, $c$ and $d$ will be more complicated of
course. At the same time, the levels coupled by an off-diagonal
matrix element will still follow the pattern revealed at $N=2$.
This means that, say, the growth of $c$ will cause a mutual
attraction of the energy levels $-3$ and $-1$, etc. Obviously, the
separate effects of attraction will compete. One even might
encounter the usual crossing of levels, not accompanied by any
instability and/or subsequent complexification of the pairs of the
levels involved. This is the reason why the ``first nontrivial"
$N=4$ model deserves a deeper analysis.

\section{Constructive analysis of
the four-by-four model\label{analysis}}

{\it A priori} we may say that the influence of the variation of
all the quadruplet of coupling constants in (\ref{prob4}) is
tractable non-numerically since the spectrum of energies coincides
with the set of roots of the secular determinant
 \be
 \det \left (\begin {array}{cccc} -3-{{E}}&0&c&b\\\noalign{\medskip}0&1-{
 {E}}&a&d\\\noalign{\medskip}-c&-a&-1
 -{{E}}&0\\\noalign{\medskip} -b&-d&0&3-{{E}}\end {array}\right )
 =0\,.
 \label{prob4f}
 \ee
The exact energies remain obtainable using closed formulae since
the corresponding secular polynomial is of the mere fourth order,
 \be
 {{{E}}}^{4}-\left (10-{a}^{2}-{b}^{2}-{c}^{2}-{d}^{2}\right )
 {{{E}}}^{2}-4\,\left ({c}^{2}-{d}^{2}\right )
 {{E}} +C(a,b,c,s)=0\,
 \label{secu4}
 \ee
where we abbreviated
 \ben
 C(a,b,c,d)=9-9\,{a}^{2}-{b}^{2}+3\,{c}^{2}+3\,{d}^{2}+{a}^{2}{b}^{2}
 +{c}^{2}{d}^{2}-2\,abcd\,.
 \een
Still, the use of the closed formulae does not facilitate our
insight in the structure of the spectrum too much as it proves
prohibitively uncomfortable. Our experience is that virtually any
alternative analytic approach to eq.~(\ref{secu4}) proves
preferable.

\subsection{Quadruple mergers of the energy levels}

We intend to describe the mechanism of a complexification of the
energies in the manner which would separate the essential and
inessential influence of the variations of the parameters. Thus,
in a formal language we shall search for the values of the matrix
elements $a$ bis $d$ at which an abrupt, qualitative change of the
spectrum could occur.

In this sense, the most interesting situation occurs at the
``points of maximal nonhermiticity"'(PMN)
at
which all the four energy levels coincide,
$E_0=E_1=E_2=E_3=z=z^{(PMN)}$. In the light of an ``up-down"
symmetry of our unperturbed spectrum $(-3,-1,1,3)$ we
may fix $z^{(PMN)}=0$. A change of this value
could only be caused by a (presumably,
perturbative) modification of our model.

Under the assumption $z=0$ our secular equation should read
$(E-z)^4=E^4=0$ so that the  quadratic term in eq.~({\ref{secu4})
must vanish,
 \be
 {a}^{2}+{c}^{2}+{b}^{2}+{d}^{2}=10\,.
 \ \ \ \ \
 \label{firsty}
 \ee
This means that all the four PMN parameters must lie on a
four-dimensional sphere with radius $\sqrt{10}$. Similarly, from
the condition of the vanishing of the linear term we deduce that
$c^2=d^2$. Finally, the condition $C(a,b,c,d)=0$ reads
 \ben
 9-{b}^{2}-9\,{a}^{2}+3\,{d}^{2}+3\,{c}^{2}
 +{c}^{2}{d}^{2}-2\,cdba+{a}^{2}{b}^{2}
 =0
 \een
and degenerates to the factorized relation
            %=============
 \be
 C(a,b,c,d)=\left (
 d^2-ab+3
 \right )^2-(b-3a)^2=
 \left (d^2-\alpha\right )\left ({d}^{2}-\beta \right
 )=0
 \label{facto}
 \ee
where $\alpha=(b+3)(a-1)$ and $\beta=(b-3)(a+1)$. This means that
at any fixed value of $d^2 >0$ we get all its solutions $(a,b)$ as
points in the $a-b$ plane which lie on the four branches of the
two hyperbolas $d^2=\alpha(a,b)$ and $d^2=\beta(a,b)$ as displayed
for illustration in Figure~1, with their two centers marked by the
bigger circles and with the two intersections marked by the small
circles (units and axes are dropped here as irrelevant).

Once we return to the former constraint (\ref{firsty}) we may
conclude that the points on the hyperbolas are spurious unless
they lie also on the centered circle with the radius
$\sqrt{10-2\,d^2}$. Hence, under our spectrum-symmetry assumption
$z=0$  there exist four PMN matrix-element solutions which induce
a ``maximal", quadruple merger of the real energy levels, in a
finite interval of values of the free parameter $d^2$ of course.
In our illustrative Figure~1 (where we choose $d^2=8/5$) we see
that and how the resulting points of the boundary
$\p {\cal D}$ in the $a-b$ plane
(denoted by symbols $C2a$, $C2b$, $C5a$ and $C5b$) emerge as
intersections of the central circle with the respective hyperbolic
segments $C2-C3$ and $C5-C6$.

\subsection{Simplified four-by-four model with $c^2=d^2$
 \label{secora} }

In terms of the abbreviations
 \ben
 A=5-{d}^{2}-\frac{1}{2}\,\left ({a}^{2}+{b}^{2}\right),
 \ \ \ \ \
 B= \left ( {d}^{2}-{a}^{}{b}^{}+3 \right )^2
 -\left({b}^{}-3\,{a}\right)^2
 \label{secu4b}
 \een
the symmetry assumption $c^2=d^2$ makes our original secular
eq.~(\ref{secu4}) simpler,
 \ben
 {{{E}}}^{4}-2\,A
 {{{E}}}^{2}
 +B=0,
 \een
and much more easily solvable by the compact formula,
 \be
 E_{\pm,\pm}=\pm \sqrt{A\pm\sqrt{A^2-B}}\,.
 \label{con}
 \ee
This means that the necessary and sufficient condition of the
reality of the energies is given by the pair of requirements
 \be
 A\geq 0
 \label{cona}
 \ee
and
 \be
 A^2\geq B \geq 0\,.
 \label{conb}
 \ee
Conditions (\ref{cona}) and (\ref{conb}) represent an
exceptionally transparent implicit definition of the
quasi-Hermiticity domain ${\cal D}$ and/or of its boundary set $\p
{\cal D}$ of {\em all} the complexification points.

Complementing the discussion presented in paragraph 3.1 above we
might notice that $B \equiv C(a,b,d,d)$ in our older notation.
This means that the two hyperbolas of Figure~1 represent precisely
the boundary curves of the domain ${\bf D}$ of validity of
condition $B \geq 0$ of eq.~(\ref{conb}). One can easily verify
that its subdomain where $d^2\leq \alpha(a,b)$ {and} $d^2\leq
\beta(a,b)$ consists of two disjoint subsubdomains ${\bf
D}{(+,A/B)}$ with the respective boundary curves $A1-A2-A3$ and
$B1-B2-B3$. Similarly, the second, single and simply connected
subdomain ${\bf D}{(-,C)}$ of ${\bf D}$ where $d^2\geq
\alpha(a,b)$ {and} $d^2\geq \beta(a,b)$ is specified by its two
pieces of boundary $C1-C2-C3$ and $C4-C5-C6$ in Figure~1.
Obviously, just the latter subdomain has a non-vanishing overlap
with the interior of the circumscribed circle (\ref{firsty}).

We can summarize that the bound-state energies of the model can
only remain real inside the latter overlap. In order to arrive at
a corresponding sufficient condition, one has to recall the last
constraint $A^2\geq B$ of eq.~(\ref{conb}). In its entirely
explicit form it reads
 \ben
 \left ( 8+a^2-b^2\right )^2 \geq 4\,d^2\,
 \left [16-(a+b)^2
 \right ]\,.
 \een
Its exhaustive discussion and geometric interpretation gets
facilitated and becomes more or less elementary in its alternative
representation
 \ben
 (2+\sigma\,\delta)^2 \geq d^2\left (4-\sigma^2
 \right )\,.
 \een
in the new, rotated coordinates $\sigma=a+b$ and $\delta=a-b$.

\section{Special case: ${\cal PT}-$symmetric band matrices
  \label{Margot} }

\subsection{Perturbative considerations}

A re-numbering of the basis  (i.e., an interchange of its second
and third element) makes the matrix in eq.~(\ref{prob4f})
equivalent (i.e., isospectral) to another Hamiltonian,
  \be
 H(a,b,c,d)=\left (\begin {array}{cccc} -3&c&0&b\\
  -c&-1&-a&0 \\
 0&a&1&d\\
 -b&0&-d&3
 \end {array}
 \right )\,.
 \label{nemarnit}
  \ee
Once the coupling of the most distant levels vanishes, $b=0$, and
once we re-install the symmetry $c=d$, we arrive at a perceivably
simpler two-parametric Hamiltonian
  \be
 H(a,c)=\left (\begin {array}{cccc} -3&c&0&0\\
  -c&-1&-a&0 \\
 0&a&1&c\\
 0&0&-c&3
 \end {array}
 \right )\,.
 \label{umarnit}
  \ee
It is particularly suitable for perturbative analysis. For
example, its one-parametric special case
  \be
 H (\alpha) = \left (
 \begin{array}{cccc}
 -3&2\alpha&0&0\\
 -2\alpha&-1&2\alpha&0\\
 0&-2\alpha&1&2\alpha\\
 0&0&-2\alpha&3
 \ea
 \right )\,
 \label{marnice}
  \ee
possesses the easily evaluated energies
 \ben
E_{\pm 1}=\pm
 \left [-6\,\alpha^2+5-2\,(5\,\alpha^4-12\,\alpha^2+4)^{1/2}
 \right ]^{1/2},
 \een
 \ben
 E_{\pm 3}=\pm
 \left [-6\,\alpha^2+5+2\,(5\,\alpha^4-12\,\alpha^2+4)^{1/2}\right
 ]^{1/2}\,.
 \een
In the regime of a small $\alpha^2$ the quickly decreasing curve
 \ben
 \left |
 E_{\pm 3}
 \right |
 =3-2\,{\alpha}^{2}-{\alpha}^{4}-{\frac {7}{6}}{\alpha}^{6}+O\left
({ \alpha}^{8}\right )
 \een
gets closer and closer to the slowly increasing curve
 \ben
 \left |
 E_{\pm 1}
 \right |
 =1+{\alpha}^{4}+{\frac{3}{2}}{\alpha}^{6}+O\left ({\alpha}^{8}\right
 ).
 \een
The energy curves finally intersect, pairwise, at a certain
critical strength,
 \ben
 \alpha^{(CS)}=\sqrt{\frac{2}{5}}, \ \ \ \ \
 E_{\pm 1}^{(CS)}=E_{\pm 3}^{(CS)}
 = \pm \sqrt{\frac{13}{5}}\sim \pm 1.612451550\,.
 \een
Beyond this boundary, i. e., at $\alpha^2>2/5$, all the
four energies become complex.

\subsection{Facilitation of the construction of the metric $\Theta$ }

There exists a clear contrast between the robust reality of the
energies resulting from a Hermitian Hamiltonian $H=H^\dagger$ and
the globally fragile character of the reality of the spectrum in
the models which are non-Hermitian and, in particular, ${\cal
PT}-$symmetric. We emphasized in section \ref{locals} that this
contrast finds a formal representation in the transition to a
nontrivial physical metric $\Theta\neq I$.

On the formal level the operator $\Theta$ may be different for
different Hamiltonians so that {\em both} the Hamiltonian $H$ {\em
and} the metric $\Theta$ may depend on certain variable
parameters. One expects, in particular, that the spectrum of $H$
ceases to be real out of the domain ${\cal D}$ of these
parameters. Of course, a necessary deeper study of all these
possibilities is much easier at finite dimensions $N$ when the
{\em linear} equation~(\ref{quasih}) determines {\em all} the
eligible metrics $\Theta$.

The straightforward linear-algebraic construction of $\Theta$
remains ambiguous. For our present, drastically simplified $N=4$
input Hamiltonians $H$ the {\em complete} solution and discussion
of the problem remains feasible. For illustration let us consider
the one-parametric model (\ref{marnice}) and solve the related
problem
 \be
 H^\dagger(\alpha)\,\Theta = \Theta\,H(\alpha)
 \label{tosolve}
 \ee
by brute force. This gives the following nontrivial
four-parametric real symmetric matrix solution
  \be
 \Theta (p,q,r,s) = \left (
 \begin{array}{cccc}
 \Theta_{11}& \Theta_{12}& r& \Theta_{14}\\
 \Theta_{12}& p& \Theta_{23}& s\\
 r& \Theta_{32}& \Theta_{33}& \Theta_{34}\\
 \Theta_{14}& s& \Theta_{34}& q
 \ea
 \right )\,
 \label{nemat}
  \ee
of the sixteen quasi-Hermiticity conditions (\ref{tosolve}). In
the solution which is routine we may employ the notation
 \ben
 \Theta_{11}=\frac{1}{6}
 \left (-9\,p+3\,q+
 10\,r+s
 \right )+\frac{1}{\alpha^2}
 \left (2\,r-s
 \right ),
 \een
 \ben
  \Theta_{14}=-\frac{(r+s
  )\alpha}{3}, \ \ \ \ \ \ \
 \Theta_{33}=\frac{1}{6}
 \left (-3\,p-3\,q+
 4\,r+s
 \right )+\frac{s}{\alpha^2}
 ,
 \een
 \ben
 \Theta_{12}=\frac{\alpha}{6}
 \left (3\,p-3\,q-
 4\,r-s
 \right )+\frac{1}{\alpha}
 \left (-2\,r+s
 \right ),
 \een
 \ben
 \Theta_{23}=\frac{\alpha}{6}
 \left (-3\,p+3\,q+
 2\,r-s
 \right )-\frac{s}{\alpha},
 \een
 \ben
 \Theta_{34}=\frac{\alpha}{6}
 \left (3\,p-3\,q-
 4\,r-s
 \right )-\frac{s}{\alpha}\,.
 \een
which specifies the unindexed matrix elements as independent
parameters.

\subsection{Construction of the surface $\p {\cal D}$
near the nonperturbative PMN regime}

In the light of our previous results, Hamiltonian $H(a,c)$ of
eq.~(\ref{umarnit}) possesses the quadruply degenerate energy
$E=E^{(PMN)}=0$ at the four PMN points with coordinates
$a=a^{(PMN)}=\pm 2$ and $c=c^{(PMN)}=\pm \sqrt{3}$. In the
vicinity of one of them (let us pick up, say, the lower left one)
we may set $a=a^{(PMN)}\,(-1+a')$ and $c=c^{(PMN)}\,(-1+c')$ with
some small measures of deviation $a'$ and $c'$.

In the zeroth order of perturbative analysis this ansatz just
reproduces the PMN solution $a'=c'=0$. On the first-order level of
precision the result  $a'=c'$ remains indeterminate. We have to
switch to an improved ansatz containing a new, auxiliary small
parameter $t$,
 \ben
 a=a^{(PMN)}\,\left [-1+t+\alpha\,t^2 +{\cal O}\left (t^3\right )
  \right ], \ \ \ \
 c=c^{(PMN)}\,\left [-1+t+\gamma\,t^2 +{\cal O}\left (t^3\right )
  \right ]\,.
  \een
Its insertion in the polynomial secular equation $\det
[H(a,c)-E]=0$ (which is of the second order in $s=E^2$) leads just
to a re-arranged version of the solutions derived in paragraph
\ref{secora}. In particular, on the second order level of
precision we obtain the following simplified version of
eq.~(\ref{cona}),
 \ben
 10\,t+\left (-5+4\,{\alpha}+6\,{\gamma}\right ){t}^{2}+O\left
 ({t}^{3 }\right )\geq 0
 \een
which only requires that our small parameter must be non-negative,
$t\geq 0$. The second half (\ref{conb}) of the implicit definition
of the quasi-Hermiticity subdomain ${\cal D}$ in the $a-c$ plane
is more informative and gives the final, comprehensive estimate
 \ben
 \gamma+\frac{8}{9}+{\cal O}(t)\geq \alpha \geq
 \gamma-\frac{1}{2}+{\cal O}(t)\,.
 \een
This formula characterizes the ``allowed" parameters $a$ and $c$
which remain compatible with the reality of the energies. Its form
is suitable for the parametric graphical plotting of the boundary
$\p {\cal D}$. The result is sampled in Figure~2 showing that in
the vicinity of the PMN matrix elements the domain ${\cal D}$ has
the shape of an extremely narrow spike. Its vertex $\left
(a^{(PMN)},c^{(PMN)} \right )$ represents the simultaneous maximum
of the size of these elements, saturating the circumscribed-sphere
inequality~(\ref{cona}) at the same time.

% \ben
% 8+9\,{\gamma}-9\,{\alpha}+O\left ({t}
% \right )\geq 0
% \een
%and
% \ben
% 1-2\,{\gamma}+2\,{\alpha}+O\left
% ({t} \right ) \geq 0\,.
% \een
%
%...

\section{Towards more-dimensional models
 \label{dodatekkectyrem}
}

A broad class of modifications of the standard harmonic
oscillators may be characterized by a certain user-friendliness of
their perturbative study. {\it A priori}, this experience may be
extended to the quasi-Hermitian models where their $N-$state
matrix Hamiltonian is just a small perturbation of the ordinary
harmonic oscillator.
Beyond this perturbative regime, unfortunately, the effects of the
non-Hermitian components become less predictable.
Firstly, in contrast to the usual textbook quantum theory where
$\Theta =I$, our present use of $\Theta \neq I$ (i.e., of a
manifest non-Hermiticity of $H$) may mean that the domain ${\cal
D}$ (where $H$ represents an observable) is finite and that many
of  the textbook perturbation-theory theorems and algorithms may
cease to be applicable~\cite{Kato}.

In particular, our present study of a specific four-state toy
model revealed that certain deeply non-perturbative mathematical
as well as physical phenomena may occur along the boundary $\p
{\cal D}$. Thus, we may expect that perturbation theory can offer
a reliable qualitative description of the most relevant
consequences of the variation of the matrix elements only in the
regime far from the boundary $\p {\cal D}$. In its vicinity, on
the contrary, perturbative considerations must be used with much
more care and in an accordingly modified form.

Several purely theoretical questions emerge near $\p {\cal D}$
also in the areas of non-quantum physics exemplified, say, by
magnetohydrodynamics \cite{Oleg}, cosmology \cite{AliWW}, crystal
optics \cite{Berry} or statistical physics \cite{Jain}. In
parallel, the points of $\p {\cal D}$ play an important role in the purely
mathematical framework of perturbation theory \cite{Calicetib} or
supersymmetric considerations \cite{jaKG}. For all these reasons
our present constructive study of the boundaries $\p {\cal D}$
may prove relevant in many different applications, after an
appropriate generalization of our schematic model if necessary.

In this context, our study of the first nontrivial $N=4$ model
offered several useful hints. We saw that our understanding and
reconstruction of the shape of the boundary $\p {\cal D}$ will
play a key role in the appropriate necessary modifications and
applications of perturbation techniques. In such a context, it is
of course unpleasant that the number of the relevant matrix
elements (i.e., of the freely variable parameters at hand) grows
very quickly with the dimension $N$ since ${\rm dim}\,{\cal D}=
entier[N^2/4]$ in general. This makes the present $N=4$ model
quite exceptional because in the very next $N=6$ model one already
has ${\rm dim}\,{\cal D}= 9$, etc.

In the purely formal setting, a sufficiently well-motivated
reduction of the number of the ``relevant" matrix elements should
be proposed in the future, therefore. The very first steps in this
direction have only been made very recently -- in ref.~\cite{egoi}
certain additional symmetries have been introduced via certain
non-Hermitian parity-type operators ${\cal P}\neq {\cal P}^\dagger
$, etc.

In the more realistic considerations the relevance of the present
model relates to the situations where some of the energy levels of
a quantum system get close to each other. A number of experimental
as well as theoretical challenges is encountered. On one side,
during a variation of parameters the so called avoided level
crossings may be observed in some nonrelativistic systems like
atomic nuclei \cite{Slavek}. On the other side, a confluence of
the two energy levels (at a point of $\p {\cal D}$) may be followed by
their subsequent complexification.

In the vicinity of a point of $\p {\cal D}$ a nontrivial innovation of
the physics of the model is often needed in its phenomenological
applications. For illustration we may recollect an electron in a
critically strong field where the single-particle Dirac equation
must necessarily be replaced by its field-theoretical extension
including many new degrees of freedom~\cite{Greiner}. In a related
brief comment \cite{Nana} we emphasized that even on the level of
the practical analyses of quantum systems using some
oversimplified phenomenological models it is not always easy to
draw the clear separation line between the avoided and unavoided level crossings. A reliable separation of the two seem
strongly model-dependent at present. All the future extension of
the scope of the quantitative analysis of the models will be
welcome, therefore.

A deeper study of the phenomenon of the complexification of the
energies to larger dimensions will be well motivated not only by
its purely mathematical appeal but also by the very pragmatic
needs of a clarification of the possible and eligible patterns of
the spectra in phenomenological models. In this sense, our present
selection of the specific illustrative ${\cal PT}-$symmetric
Hamiltonians $H$ in a certain ``first nontrivial" matrix form may
be perceived as a natural starting point of such an effort.

%\section*{Acknowledgement}
%
%Supported by GA\v{C}R, grant Nr. 202/07/1307.
%

%%Institutional Research Plan AV0Z10480505 and by
%%the M\v{S}MT ``Doppler Institute" project Nr. LC06002.
%
%%\end{document}

\section*{Acknowledgement}

Supported by GA\v{C}R, grant Nr. 202/07/1307.

%Institutional Research Plan AV0Z10480505 and by
%the M\v{S}MT ``Doppler Institute" project Nr. LC06002.

%\end{document}

\section*{Figure captions}

\subsection*{Figure 1.
The centered circle (\ref{firsty}) and the two hyperbolas
$C(a,b,d,d)= 0$ with the respective centers at $(a,b)=(-1,3)$ and
$(a,b)=(1,-3)$ (marked by medium circles) in $a-b$ plane at
$c^2=d^2=1.6$}

\subsection*{Figure 2. Spiked shape of the physical domain
 ${\cal D}(a,c)$ near its lower left corner}

 
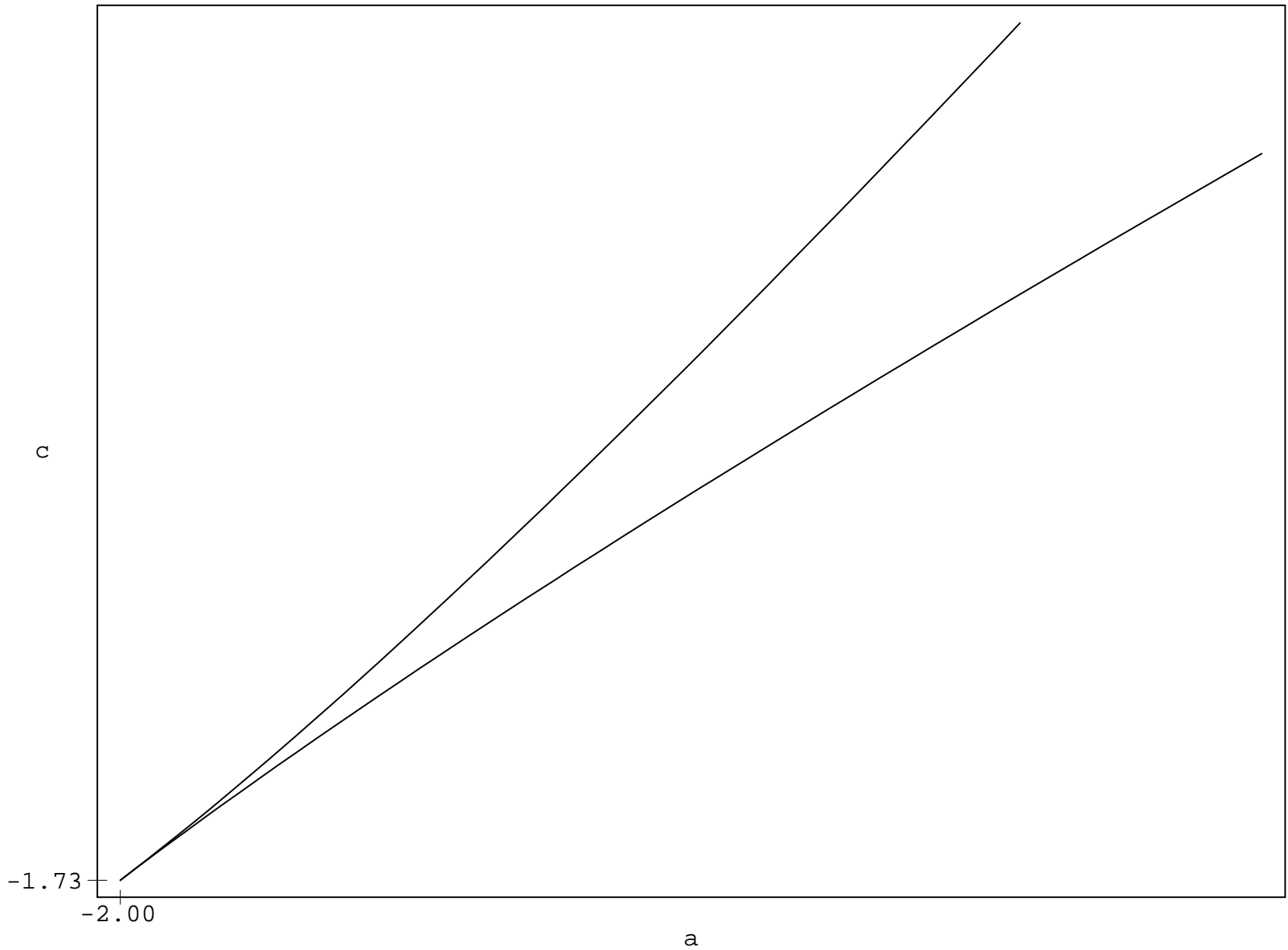
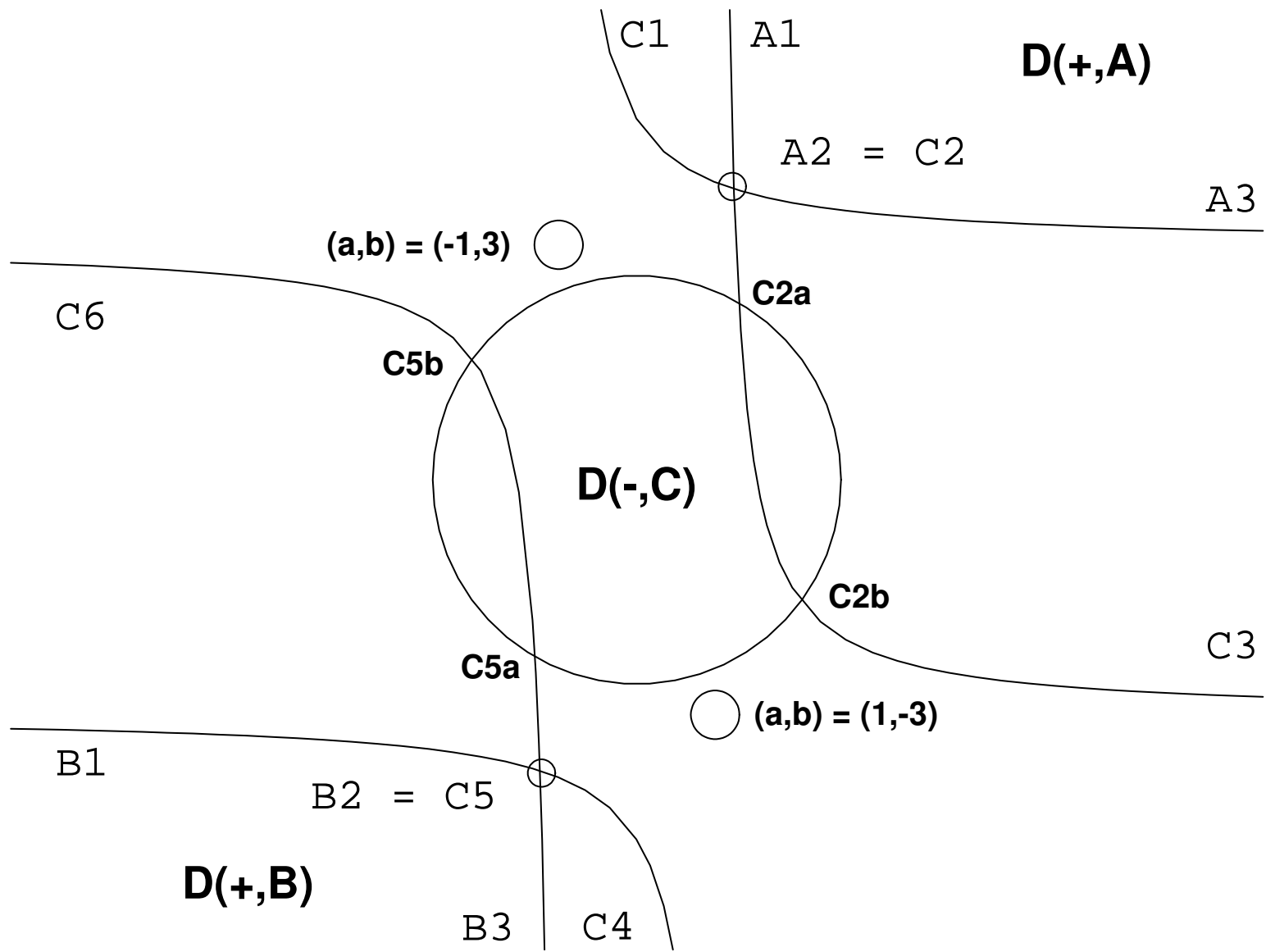
\end{document}